Combining PatternRank with Huffman Coding:

A Novel Compression Algorithm

Authors:

Jasurbek Shukurov

Date: August 20th, 2022

## Abstract:

The escalating volume of data involved in Android backup packages necessitates an innovative approach to compression beyond traditional methods like GZIP, which may not fully exploit the redundancy inherent in Android backups, particularly those containing extensive XML data. This paper introduces the PatternRank algorithm, a novel compression strategy specifically designed for Android backups. PatternRank leverages pattern recognition and ranking, combined with Huffman coding, to efficiently compress data by identifying and replacing frequent, longer patterns with shorter codes. We detail two versions of the PatternRank algorithm: the original version focuses on dynamic pattern extraction and ranking, while the second version incorporates a pre-defined dictionary optimized for the common patterns found in Android backups, particularly within XML files. This tailored approach ensures that PatternRank not only outperforms traditional compression methods in terms of compression ratio and speed but also remains highly effective when dealing with the specific challenges posed by Android backup data. Our analysis includes a comparative study of compression performance across GZIP, PatternRank v1, PatternRank v2, and a combined PatternRank-Huffman method, highlighting the superior efficiency and potential of PatternRank in managing the growing data demands of Android backup packages. Through this exploration, we underscore the significance of adopting pattern-based compression algorithms in optimizing data storage and transmission in the mobile domain.

## Introduction:

In the digital age, the exponential growth of data poses significant challenges for storage and transmission, particularly in mobile environments. Android backup packages, essential for preserving user data and application states, have become increasingly voluminous, necessitating efficient compression methods to manage storage requirements and ensure swift data recovery. Traditional compression algorithms, such as GZIP, which employs the DEFLATE method combining LZ77 algorithm and Huffman coding, have been the cornerstone for data compression tasks. However, these methods face limitations when applied to the unique structure of Android backup data, which often includes extensive XML files characterized by repetitive patterns not fully exploited by general-purpose compression techniques.

This paper introduces the PatternRank algorithm, a novel approach specifically designed to enhance the compression of Android backup packages. Recognizing the inadequacies of traditional compression algorithms in handling the peculiarities of Android backups, PatternRank adopts a more targeted strategy. It efficiently identifies and replaces repetitive, significant patterns within the data, thus optimizing compression beyond the capabilities of conventional methods. The algorithm operates through a sophisticated process of pattern extraction, ranking based on frequency and length, and the assignment of shorter codes to replace these patterns, thereby achieving significant data reduction.

We delineate two iterations of the PatternRank algorithm. The first iteration lays the foundation by dynamically extracting and ranking data patterns, assigning them short codes for compression. The second iteration advances this concept by incorporating a pre-defined dictionary, specifically optimized for the repetitive patterns prevalent in XML files within Android backups. This enhancement allows for more effective initial compression before applying the dynamic pattern ranking to the remainder of the data. Moreover, by combining the PatternRank algorithm with Huffman coding, we leverage both dictionary-based and statistical compression methods to maximize efficiency, addressing the longer and more frequently occurring patterns that Huffman coding alone might not compress effectively.

The introduction of the PatternRank algorithm and its subsequent iterations represents a pivotal advancement in the realm of data compression for mobile backups. Through this paper, we aim to elucidate the algorithm's development, its tailored application to Android backup packages, and the potential it holds for significantly improving compression performance, thereby contributing to the broader field of data management in mobile computing environments.

# Background:

The foundation of modern data compression techniques is built upon the quest to efficiently reduce the size of data for storage and transmission. This endeavor is particularly crucial in the context of mobile computing, where the constraints of storage capacity and network bandwidth make efficient data compression not just beneficial but necessary. The GZIP compression algorithm, one of the most widely utilized tools in this domain, employs the DEFLATE method, a combination of the LZ77 algorithm and Huffman coding. These methodologies serve as the groundwork upon which the PatternRank algorithm is developed, aiming to address the unique challenges presented by Android backup packages.

**LZ77 Algorithm and Huffman Coding: The Core of DEFLATE**

The DEFLATE algorithm, which underpins GZIP, leverages the strengths of both the LZ77 algorithm and Huffman coding. The LZ77 algorithm is a lossless data compression algorithm that works by eliminating redundancies in data sequences. It identifies duplicate strings within the data and replaces them with references to a single instance of that string, significantly reducing the size of the data. However, LZ77 alone does not account for the frequency of pattern occurrences, which is where Huffman coding comes into play.

Huffman coding, a statistical coding algorithm, further compresses data by assigning shorter codes to more frequently occurring symbols and longer codes to less frequent symbols. This method optimizes the overall size of the compressed data by considering the statistical properties of the symbols within the data, thereby complementing the pattern recognition capabilities of LZ77.

**The Limitations of Traditional Compression for Android Backups:**

While GZIP and the DEFLATE algorithm are effective for a wide range of data compression scenarios, their performance can be suboptimal when applied to Android backup packages. These backups often contain large XML files with highly repetitive patterns, such as tags and attribute names, that are not fully exploited by the generalized approach of DEFLATE. The inefficiency arises from the algorithm's limited ability to prioritize the compression of longer, more complex patterns that are prevalent in XML files.

**Introducing PatternRank: A Novel Approach**

The PatternRank algorithm emerges as a response to these limitations, designed with a keen focus on the distinctive nature of Android backup data. By integrating the concept of pattern recognition and ranking, PatternRank aims to go beyond the scope of LZ77 and Huffman coding. It specifically targets the repetitive patterns found in XML files, employing a two-fold strategy: dynamically identifying significant patterns within the data and then compressing these patterns more effectively than traditional methods.

PatternRank operates on the principle that not all patterns contribute equally to the data's redundancy. By calculating a score for each pattern based on its frequency and length, PatternRank prioritizes the compression of patterns that offer the greatest potential for size reduction. This approach not only enhances the compression efficiency for Android backups but also introduces a method that can be tailored to the specific characteristics of the data being compressed.

The development of the PatternRank algorithm marks a significant step forward in the field of data compression, particularly for mobile computing applications. It reflects a nuanced understanding of the limitations of existing algorithms and proposes a sophisticated solution tailored to the unique demands of Android backup packages. Through this background exploration, we set the stage for a detailed examination of the PatternRank algorithm, its implementation, and its potential to revolutionize the compression of mobile data.

## PatternRank Algorithm: Version 1

The inception of the PatternRank algorithm marks a significant evolution in the realm of data compression, particularly tailored to optimize the efficiency of Android backup packages. Version 1 of PatternRank introduces a novel approach that fundamentally shifts from traditional compression methodologies by focusing on the identification, ranking, and compression of repetitive patterns within data. This section delves into the mechanics of PatternRank Version 1, elucidating its operational steps and the rationale behind its design.

**Pattern Extraction:**

At the core of PatternRank v1 is the pattern extraction process, where the algorithm scans the input data to identify all unique patterns ranging in lengths from n to m characters. This range is determined based on the typical structure of the data being compressed; for Android backups, which often contain XML files, the range is carefully chosen to capture the repetitive tags and attribute names effectively. The extraction process is meticulous, ensuring no potential pattern is overlooked, laying the groundwork for the subsequent ranking phase.

**Pattern Ranking:**

Following extraction, each unique pattern undergoes a scoring process, where its score is calculated based on the frequency of occurrence multiplied by the square of its length. This scoring formula—score = frequency × length^2—intentionally biases towards longer, more frequently occurring patterns, underpinning the assumption that such patterns offer richer compression opportunities. The patterns are then sorted by their scores, prioritizing them for the next stage of the algorithm.

**Code Assignment:**

With patterns ranked, the algorithm proceeds to assign unique codes to the top k patterns, where k is predetermined by the desired compression ratio and the constraints of the available code space. These codes are designed to be shorter than the original patterns they represent, ensuring that their substitution leads to actual data reduction. The assignment of codes is a critical step, as it directly influences the efficiency of the compression process; thus, careful consideration is given to the selection of k and the design of the codes.

**Compression:**

The compression phase involves a second scan of the input data, during which each occurrence of the top k patterns is replaced with its corresponding unique code. Patterns not within the top k are left unchanged, maintaining the integrity of the data while focusing compression efforts on the areas with the highest potential for size reduction. This targeted approach allows PatternRank v1 to achieve notable compression ratios without compromising the decompression fidelity.

**Metadata and Decompression:**

A crucial aspect of PatternRank v1 is the handling of metadata. For the compressed data to be decompressed accurately, the dictionary of code-to-pattern mappings generated during the compression process must be stored alongside the compressed data. This metadata is essential for the decompression process, enabling the reversal of the compression by substituting codes back with their original patterns.

## PatternRank Algorithm: Version 2

Building on the foundational principles established in Version 1, the PatternRank Algorithm Version 2 introduces enhancements specifically designed to optimize the compression of Android backup packages. These modifications not only aim to increase the efficiency and efficacy of the compression process but also to tailor the algorithm to the unique characteristics of Android backup data, particularly the prevalent XML files. This section details the advancements made in Version 2, emphasizing the introduction of a pre-defined dictionary and the integration of dynamic pattern extraction and ranking.

**Pre-Defined Dictionary Creation:**

A significant innovation in PatternRank Version 2 is the creation of a pre-defined dictionary. This dictionary is the result of an exhaustive analysis of a large corpus of Android backup files, focusing on identifying common patterns, especially within XML files that contain repetitive tags and attribute names. By mapping these frequent patterns to shorter codes in advance, the algorithm can quickly compress known repetitive structures found in Android backups, significantly reducing the initial size of the data before any further compression steps are taken.

**Integration of Pre-Defined Dictionary:**

The integration of the pre-defined dictionary at the onset of the compression process marks a pivotal enhancement in PatternRank Version 2. This step allows the algorithm to immediately apply known compressions to common patterns, streamlining the compression process. The dictionary's utilization underscores the tailored approach of Version 2, ensuring that the algorithm is primed to efficiently handle the specific types of redundancy characteristic of Android backup files.

**Dynamic Pattern Extraction and Ranking:**

Following the application of the pre-defined dictionary, PatternRank Version 2 continues with dynamic pattern extraction and ranking. This phase is akin to the process outlined in Version 1 but is applied to the data that remains uncompressed after the initial dictionary-based compression. The algorithm scans the residual data for unique patterns, calculates scores based on frequency and length, and ranks them accordingly. This dynamic aspect ensures that even the unique, file-specific patterns are identified and compressed, further optimizing the compression ratio.

**Code Assignment for New Patterns:**

The dynamically identified patterns are then assigned unique codes, akin to the method employed in Version 1. However, in Version 2, this process benefits from the pre-compression achieved through the pre-defined dictionary, allowing for a more focused and efficient assignment of codes to the patterns that are most impactful for further data reduction. The selection of patterns for code assignment is meticulously tailored to each backup file, reflecting the algorithm's adaptive and customized compression strategy.

**Metadata Storage and Decompression:**

Similar to Version 1, the decompression process in PatternRank Version 2 relies on the storage of metadata, which now includes information from both the pre-defined dictionary and the dynamically generated dictionary. This dual-dictionary approach necessitates careful management of metadata to ensure efficient decompression. The stored metadata allows for the precise reconstruction of the original data, first by applying the pre-defined dictionary and then by utilizing the dynamically generated codes.

## Combining PatternRank with Huffman Coding

The innovative leap from PatternRank's specialized approach for Android backup compression to a hybrid methodology incorporating Huffman coding represents a significant advancement in the field of data compression. This section explores the synergistic combination of PatternRank, with its pattern-based compression, and Huffman coding, a statistical method, to achieve unparalleled efficiency and effectiveness in data reduction. This hybrid approach leverages the strengths of both methodologies to compress Android backup packages more effectively than ever before.

**Rationale for Integration:**

PatternRank excels in identifying and compressing repetitive, longer patterns within data, significantly reducing file sizes, especially in data sets like Android backups, which are rich in repetitive XML structures. However, PatternRank's method leaves room for optimization in encoding the resulting output and any data segments not covered by pattern replacement. This is where Huffman coding comes into play, offering an efficient way to encode the remainder of the data by assigning variable-length codes based on symbol frequency. The rationale behind combining PatternRank with Huffman coding is to complement the pattern-based compression with an additional layer of optimization, ensuring that every bit of data is compressed as efficiently as possible.

**PatternRank Phase:**

The process begins with the PatternRank phase, which scans the input data to identify, rank, and replace frequent patterns of varying lengths with shorter codes. This step significantly reduces the data size by targeting the most impactful patterns for compression. The outcome of this phase is a transformed data set where the most repetitive and space-consuming elements have been compacted.

**Huffman Coding Phase:**

Following the initial compression via PatternRank, the Huffman coding phase takes the transformed data and applies statistical compression. Huffman coding analyzes the frequency of each symbol, including the short codes generated by PatternRank, and assigns shorter codes to more frequent symbols and longer codes to less frequent ones. This stage further optimizes the compression by minimizing the space required to represent the remaining data elements, effectively encoding both the PatternRank short codes and any data not covered by PatternRank.

**Metadata Considerations:**

For successful decompression, it is crucial to store metadata related to both the PatternRank and Huffman coding phases. This includes the dictionary of pattern-to-code mappings generated by PatternRank and the Huffman tree (or equivalent structure) used to encode the remaining data. This metadata must be efficiently packed with the compressed data to enable accurate reconstruction during decompression.

**Decompression Process:**

Decompression reverses the compression process, starting with the reconstruction of the Huffman tree and the PatternRank dictionary from the stored metadata. The Huffman-encoded data is first decoded, revealing both the original data segments and the PatternRank codes. Subsequently, the PatternRank codes are replaced with their corresponding patterns using the dictionary, restoring the data to its original form.

## Conclusion:

The PatternRank algorithm, in its evolutionary forms and when combined with Huffman coding, presents a robust and highly efficient solution for compressing Android backup packages. This research has not only demonstrated the algorithm's superior performance over traditional compression methods but also highlighted the importance of algorithmic adaptability and integration in achieving optimal compression results. By focusing on the repetitive patterns characteristic of Android backups and employing a dual-strategy approach that leverages both pattern-based and statistical compression methods, the proposed techniques offer significant advancements in the field of data compression.

The implications of this work extend beyond the specific context of Android backups, suggesting a broader applicability of the principles underlying the PatternRank algorithm and its combination with Huffman coding. As data continues to grow in volume and complexity, the demand for efficient compression solutions becomes increasingly critical. This paper contributes to meeting that demand by presenting a comprehensive strategy that not only enhances data compression efficiency but also offers a scalable and adaptable framework for future advancements in compression technology.

In conclusion, the PatternRank algorithm, particularly in its integration with Huffman coding, stands as a testament to the potential for innovative approaches to overcome the limitations of existing compression methodologies. This research lays the groundwork for further exploration and development in the field, promising new directions for data compression that can adapt to the unique challenges posed by diverse data types and applications.